\documentclass[twocolumn,showpacs,preprintnumbers,amsmath,amssymb]{revtex4}

\usepackage{graphicx}
\usepackage{dcolumn}
\usepackage{bm}

\begin{document}



\title{Searching for vector bileptons

in polarized Bhabha scattering

}

\author{B. Meirose}
\email{bmeirose@cern.ch}
\affiliation{
Physikalisches Institut \\
Albert-Ludwigs-Universit\"{a}t Freiburg \\
Hermann-Herder-Str. 3 79104 Freiburg i. Br., Germany
}%

\author{A. J. Ramalho}%
\email{ramalho@if.ufrj.br}


\affiliation{%
Instituto de F\'{\i}sica \\
Universidade Federal do Rio de Janeiro \\
Cx.Postal 68528, 21945-970 Rio de Janeiro RJ, Brazil
}%

\date{\today}

\begin{abstract}
In this paper we analyze the effects of virtual vector bileptons in polarized
Bhabha scattering at the energies of the future linear colliders.
In order to make the calculations of the differential cross sections more
realistic, important beam effects such as initial state radiation,
beamstrahlung, beam energy and polarization spreads are accounted for. The
finite resolution of a typical electromagnetic calorimeter planned for the
new linear colliders is also considered in the simulation. The 95\%
confidence level limits for bilepton masses in 331 models are evaluated.
\end{abstract}


\maketitle

\section{\label{sec:level1}Introduction:\protect\\
} Although the standard model (SM) explains all current experimental data,
it is commonly believed that it is not the final answer. Many questions have
been left open, so many theorists believe that there should be some new
physics lurking at the TeV scale. Many models have been proposed to explain
these unanswered questions. Examples are supersymmetry, composite models and
grand unification theories. All extended models have in common the prediction
of new particles. This paper concerns nonstandard particles known as
bileptons \cite{CUYDAV} - bosons which couple to two leptons and which carry
two units of lepton number. They are present in several models, such as
technicolor, left-right symmetric models, and in the grand unification
schemes. Heavy gauge bileptons appear in extended gauge models
where the electroweak group is imbedded in a larger group. The so-called
 331 models, based on the gauge group $SU(3)_C \otimes SU(2)_L \otimes
U(1)_Y$, fall in this category.  In this
paper we discuss how to search for signals of an off-mass-shell,
doubly-charged vector bilepton in polarized Bhabha scattering, at
the new linear collider energies. Most of the analysis is made in the
framework of 331 gauge models \cite{FRAPLE}, but the predictions of an
$SU(15)$ model \cite{SUFI} are also presented, and may be taken as a measure
of the model-dependence of our results. The detection of such indirect effects
of a vector bilepton would be strong evidence of new physics. Should any
signal of a vector bilepton be detected in a high energy collider, it would
be important to settle the question of the correct underlying model.

\par
For realistic comparison with experiment, we consider important beam effects such as (i) initial-state radiation, beamstrahlung and beam energy spread; (ii) longitudinal and transverse polarization of the colliding beams, which are expected to be available in the next generation of linear colliders; (iii) Gaussian smearing of the four-momenta of the final-state leptons, simulating the uncertainties in the energy measurements in the electromagnetic calorimeters. In section II we give a brief review of the $331$ minimal model \cite{PPT}.
Section III describes in detail the numerical
simulation of the Bhabha events. In section IV we analyze several observables such as angular distributions and asymmetries, with the purpose to look for indirect vector bileptons signals. In section V we establish bounds on the couplings and masses of these bileptons at the $95\%$ confidence level, from the angular distribution of the final-state leptons. As we are working in the context of the minimal 331 version, one can also verify the validity of the relations between the masses of the vector bileptons and new neutral gauge bosons, which are connected to the Higgs structure of the $331$ models.

\section{Review of the $331$ minimal model}
\par
The fermionic content of the  minimal $331$ model arranges the ordinary leptons in $SU(3)_L$ antitriplets, two generations of quarks in triplets and the third generation in an antitriplet. The anomaly cancellation takes places only when all three families of quarks and leptons are taken together. The model also predicts a new neutral gauge
boson $Z^{\prime}$ and four vector bileptons $Y^{\pm}$ and $Y^{\pm
\pm}$, which acquire masses after spontaneous symmetry breaking (SSB): first $\mathrm{SU(3)_L}\times \mathrm{U(1)_Y}$ breaks down to $\mathrm{SU(2)_L}\times \mathrm{U(1)_Y}$. This can be accomplished with only one $\mathrm{SU(3)_L}$ scalar triplet; the next stage of SSB, $\mathrm{SU(2)_L}\times \mathrm{U(1)_Y}\to \mathrm{U(1)_e}$,
requires an additional scalar $\mathrm{SU(3)_L}$ triplet. In order to provide quarks and leptons with acceptable masses, however, two additional triplets and a sextet would be necessary for symmetry breaking with a minimal Higgs structure. In this minimal version of the $331$ model, the mass $M_Y$ of the doubly-charged bilepton $Y^{++}$  and the mass $M_{Z^\prime}$ of the neutral vector boson $Z^\prime$ are correlated parameters given by the expression:

$$ {M_Y \over M_{Z^\prime}} = {\sqrt{3(1-4\sin^2\theta_W)} \over 2\cos
\theta_W }$$ This relation no longer holds for the $SU(15)$ or any other model with different Higgs structure. Presently, the most useful lower bounds on the vector bilepton mass can be derived from fermion pair production at LEP and lepton-flavor violating charged lepton decays \cite{TUL}, $M_Y > 740 GeV$, and from  muonium-antimuonium conversion \cite{WILL}, $M_Y > 850 GeV$. While the former is less stringent, it does not depend on the assumption that the bilepton coupling is flavor diagonal. The data collected by the CDF Collaboration at the Fermilab Tevatron exclude 331 $Z^\prime$ masses below $920$\; $GeV$ \cite{GUTI}.
\par
It has been argued \cite{DIAS1} that the minimal $331$ model described above can not be analyzed with perturbation theory around an energy scale $\mu$, at which the ratio $g_X^2/g_L^2 = sin^2\theta_W(\mu)/(1 - 4 sin^2\theta_W(\mu))$, where $g_X$ and $g_L$ denote the $U(1)_X$ and $SU(3)_L$ gauge group coupling constants, develops a Landau-like pole. The scale $\mu$ was found to be of the order of $4 TeV$ for the minimal $331$ model. Solutions \cite{DIAS2} have been put forward to circunvent this possible nonperturbative nature of some $331$ models at the TeV energy scale. In particular, by adding three octets of vector leptons to the particle content, it is possible to make $sin^2\theta_W(\mu)$ decrease with the increase of the energy scale $\mu$, as far as the light states of the standard model are concerned, and thus the perturbative regime is restored for the TeV energy scale.

\section{Simulation}
\par
  When a charged particle is accelerated to energies much superior than its own rest energy, the probability of photon emission is very high. The so called initial-state radiation (ISR) is the emission of photons by the incoming electrons and positrons due this acceleration. ISR is a dominant QED correction in leptonic scattering. Using the structure function approach discussed in \cite{SKR}, ISR effects were taken into account in this simulation. We neglect electroweak loop corrections throughout the calculations. For the energies planned for the future linear colliders, the first-order weak loop-corrections to the Bhabha angular distribution \cite{BDH} are at most of the order of $10\%$ of the corresponding distribution at Born level, for wide-angle scattering. The remaining weak loop-corrections are smaller. We expect that the magnitude of weak loop-corrections in the extended models we are studying be similar to that of the standard model. Moreover, to some extent these
 corrections should cancel out in the asymmetry ratios, and therefore their effect on the observables discussed in the next section should be small. In the quest for higher luminosities at the linear colliders, very dense incoming beams, with transverse dimensions of the order of a few dozen nm, will be needed. The result is a large charge density, leading to very strong electromagnetic fields. For this reason, the particles in a colliding bunch suffer considerable transverse acceleration, which gives rise to the emission of synchrotron radiation, the so-called beamstrahlung. The effective energy available for the reaction is then smaller than the nominal value. Machines with large luminosity per bunch crossing produce more beamstrahlung. Thus, the average energy loss of a positron or electron by beamstrahlung depends on the design parameters of the accelerator. For some designs the beamstrahlung energy loss may even reach $30\%$  of the available energy \cite{CLIC}. Using the approach discussed on ref. \cite{PESK}  we obtained the beamstrahlung structure function, starting from an energy-dependent set of NLC design parameters \cite{BEAM}. Convoluting the ISR and beamstrahlung emissions spectra we obtained the distribution which was used to compute the required differential cross sections. Our simulations also included a Gaussian-distributed beam energy spread, with a width of $1\%$ of the nominal beam energy.
\par
As we are working mostly in the framework of the $331$ minimal model, two new Feynman diagrams have to be added to the standard diagrams for Bhabha scattering: a s-channel exchange of a $Z^\prime$ boson and a u-channel
exchange of a doubly-charged bilepton. As a partial check of our calculation, we "switched off" the beam polarizations, ISR and beamstrahlung effects and the $Z^\prime$ exchanges as well,
and  verified that we indeed reproduce the results of the trace calculation of ref. \cite{FRIZ}.
Likewise, we cross-checked our calculations with those of ref. \cite{OLS},
which were carried out in the framework of the standard model, but
with arbitrary beam polarization.
 Using Monte Carlo techniques we calculated the differential cross sections
with the simulated events selected according to the following set of cuts : (i) the final-state electrons and positrons were required to be produced within the angular range $\vert cos\theta_i \vert < 0.95$, where $\theta_i$ stands for the polar angle of the final-state lepton with respect to the direction of the incoming electron beam; (ii) all events in
which the acollinearity angle $\zeta$ of the final-state $e^+-e^-$
three-momenta did not pass the cut $\zeta < 10^\circ$ were
rejected; (iii) the ratio of the effective center-of-mass energy
to the nominal center-of-mass energy for any acceptable event was
required to be greater than $0.9$. We considered a $500$
$fb^{-1}$ integrated luminosity, which could be achieved in several years of machine operation , considering a maximum center-of-mass energy of $1$ $TeV$.
 Finally, we simulated the finite resolution of the NLC
electromagnetic calorimeters by Gaussian-smearing the four-momenta
of the produced electrons and positrons \cite{SMEAR}. The directions of the lepton three-momenta were smeared in a cone around the corresponding original directions, whose half-angle is a Gaussian with half-width equal
to $10$ $mrad$, whereas the energies of the final-state leptons were distributed as a Gaussian with half-width $\Delta E$ of the form $\Delta E/E = 10\%/\sqrt{E} \oplus 1\%$.

\par

     As the experiments at SLC have shown, beam polarization is an important tool in the search for new physics and for precision measurements in particle physics. Electron and positron beams with high degrees of polarization will play a significant role at the new linear
colliders \cite{MOOR}. Although lepton polarization has the inconvenience of being accompanied by considerable beam intensity loss, the new linear colliders will compensate by achieving very high luminosities.
Polarized lepton beams can reduce backgrounds and increase of the sensitivity of spin-dependent observables to nonstandard phenomena.

\section{Observables}

In our numerical calculations, the longitudinal (transversal) beam polarizations  were taken to be $P_L = 90\%$ $(P_T = 90\%)$ for electrons and $P_L = 60\%$ $(P_T = 60\%)$ for positrons, with an uncertainty given by $\Delta P_L/P_L = 0.5\%$  $(\Delta P_T/P_T = 0.5\%)$. Firstly, we only considered longitudinal polarization of the beams.

\begin{figure}
\graphicspath{{c:/BERNHARD/Bhabha/}}
\rotatebox{-90}{\scalebox{0.35}{\includegraphics{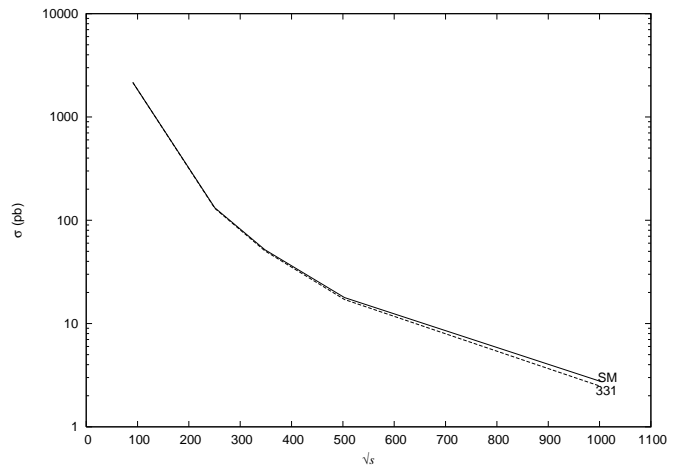}}}
\caption{\label{fig1}Total cross section as a function of the
center-of-mass energy $\sqrt{s}$, for $M_Y = 1.2 \, TeV \,$ (331);
solid line (SM) represents the standard model cross section.}
\end{figure}

  Fig.\ref{fig1} shows the dependence of the total cross section of the $331$ minimal model on the center-of-mass energy, for an input bilepton mass $M_Y = 1.2 TeV$. For acceptable values of the bilepton mass, the deviation from the  standard model cross section is small over the energy range expected for the next generation of linear colliders. This was also found to be true for the $SU(15)$ model.

\begin{figure}
\rotatebox{-90}{\scalebox{0.35}{\includegraphics{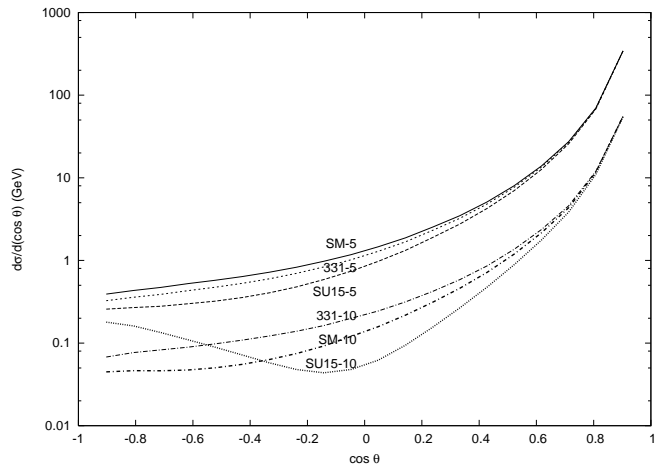}}}
\caption{\label{fig2} Angular distribution of the final-state
electrons; Curves (SM-5) and (SM-10) show the angular spectra predicted
by the standard model at $\sqrt{s} = 500 GeV$ and $\sqrt{s} = 1
TeV$ respectively, while (331-5) and (331-10) display the corresponding
angular distributions in the minimal {331} model. }
\end{figure}

Next we looked at the angular distributions in the minimal $331$ model at $\sqrt{s} = 500 GeV$ and $\sqrt{s} = 1 TeV$, which are displayed in Fig.\ref{fig2} with the corresponding curves for the standard model and $SU(15)$ shown for comparison. All curves indicate that the final-state electrons are emitted mostly in small angles (with respect to the incident electron beam), but significant differences among the angular distributions still occur over the remaining angular range. These deviations from the SM predictions could be helpful to establish the correct underlying electroweak model, should any signal of a vector bilepton be detected in a high energy collider. In a previous work \cite{MEIROSE}, a similar pattern was found for the sensitivity of M\o ller scattering total cross section and angular distribution to the coupling of vector bileptons.

\begin{figure}
\rotatebox{-90}{\scalebox{0.35}{\includegraphics{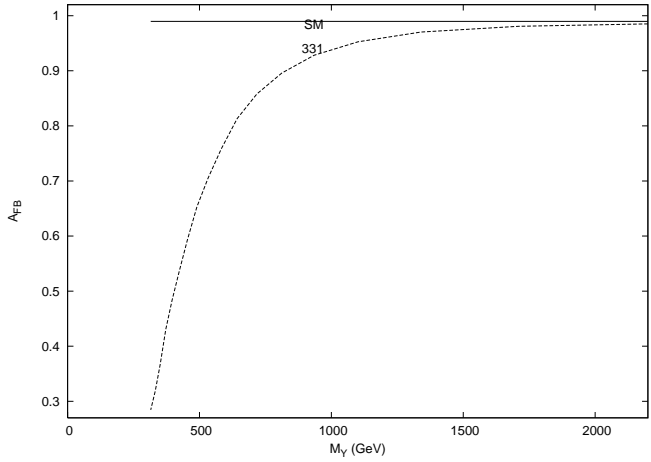}}}
\caption{\label{fig3} Forward-backward asymmetry $A_{FB}$ for
several input masses $M_Y$, at $\sqrt{s} = 1 TeV$, according to
the minimal $331$ model (331). Curve (SM) shows the standard model
prediction}
\end{figure}

As the angular distribution for 331 and standard models are strongly asymmetric, a typical non-identical particle final state behavior, the integrated forward-backward asymmetry is large:

\begin{eqnarray}
A_{FB}=\frac{\int_{-0.95}^{0} dz \frac{d \sigma}{dz}
-\int_{0}^{0.95} dz \frac{d \sigma}{dz}}{\int_{-0.95}^{0} dz \frac{d \sigma}{dz}
+\int_{0}^{0.95} dz \frac{d \sigma}{dz}}
\end{eqnarray}
where $z$ stands for the cosine of the angle $\theta$ between the outgoing and incoming electrons. This is displayed in Fig.\ref{fig3}, where $A_{FB}$ is plotted for several bilepton masses at an energy $\sqrt{s} = 500 GeV$. As expected the 331 $A_{FB}$ value approaches the standard model $A_{FB}$ value as the bilepton mass increases.

Also, we analyze the discovery potential of spin asymmetries.

Using the polarization-dependent angular distributions, we define the following asymmetries:
\begin{widetext}
\begin{equation}
A_1(cos\theta) = \frac{d\sigma(-\vert P_L^a \vert, -\vert P_L^b \vert)
+
                    d\sigma(-\vert P_L^a \vert, \vert P_L^b \vert)  -
                    d\sigma(\vert P_L^a \vert, -\vert P_L^b \vert)  -
                    d\sigma(\vert P_L^a \vert, \vert P_L^b \vert)}
                   {d\sigma(-\vert P_L^a \vert, -\vert P_L^b \vert) +
                    d\sigma(-\vert P_L^a \vert, \vert P_L^b \vert)  +
                    d\sigma(\vert P_L^a \vert, -\vert P_L^b \vert)  +
                    d\sigma(\vert P_L^a \vert, \vert P_L^b \vert)  }
\end{equation}
\end{widetext}
\begin{equation}
A_2(cos\theta) = \frac{d\sigma(-\vert P_L^a \vert, -\vert P_L^b \vert)
- d\sigma(\vert P_L^a \vert, \vert P_L^b \vert)} {d\sigma(-\vert P_L^a
\vert, -\vert P_L^b \vert) + d\sigma(\vert P_L^a \vert, \vert P_L^b
\vert)}
\end{equation}

\begin{equation}
A_3(cos\theta) = \frac{d\sigma(-\vert P_L^a \vert, \vert P_L^b \vert)
- d\sigma(0,0) }{d\sigma(-\vert P_L^a \vert, \vert P_L^b \vert)  +
d\sigma(0,0) }
\end{equation}

\begin{figure}
\rotatebox{-90}{\scalebox{0.35}{\includegraphics{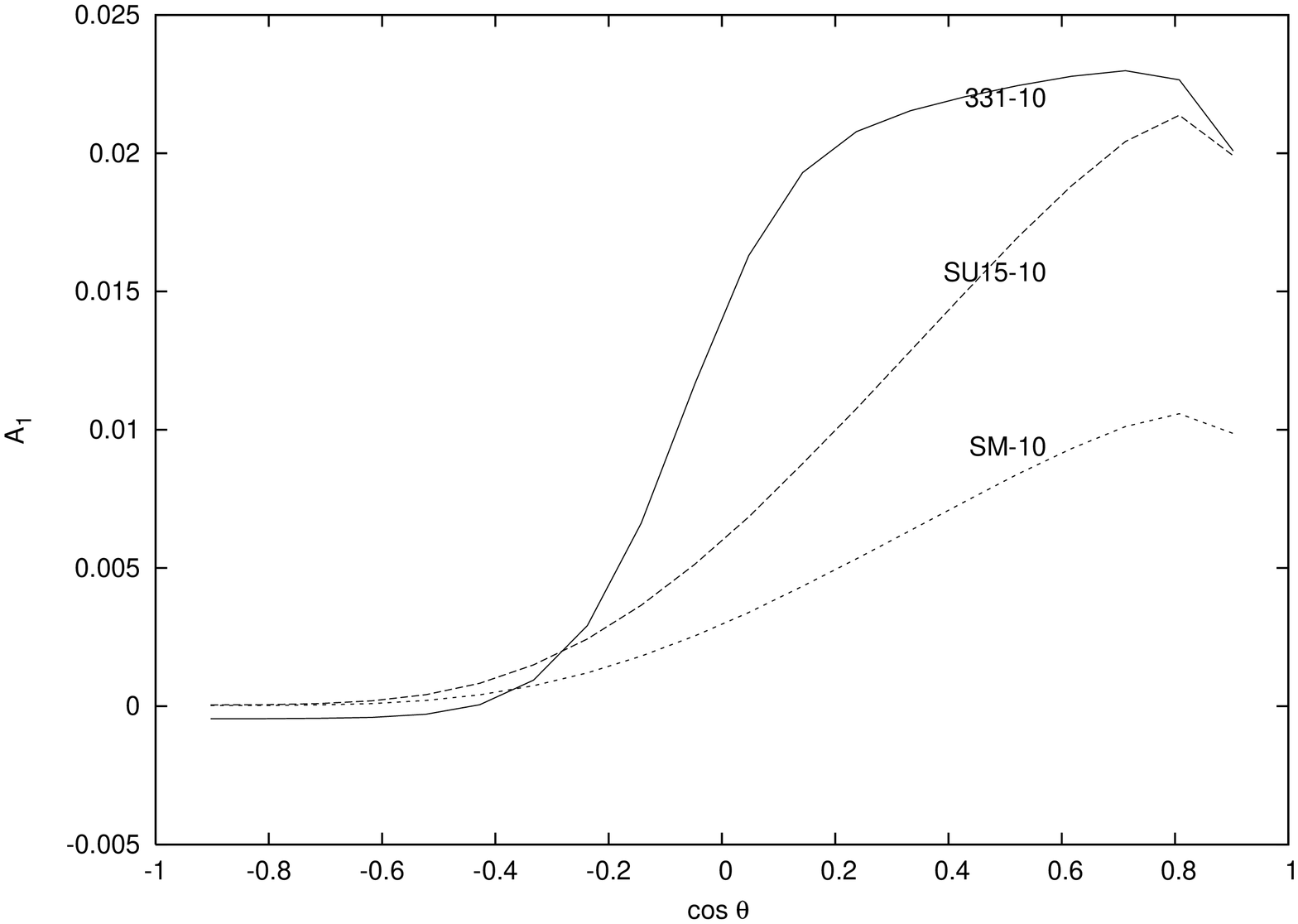}}}
\caption{\label{fig4} Polar angle dependence of spin asymmetry
$A_1(cos\theta)$ at $\sqrt{s} = 1 TeV$; Curve (SM-10) shows the
standard model predictions for $A_1(cos\theta)$, while (331-10)
represents the corresponding expectations for the minimal {331}
model, for a mass $M_Y = 1.2 TeV$. Curve (SU15-10) shows the prediction
for the SU(15) model.}
\end{figure}

\begin{figure}
\rotatebox{-90}{\scalebox{0.35}{\includegraphics{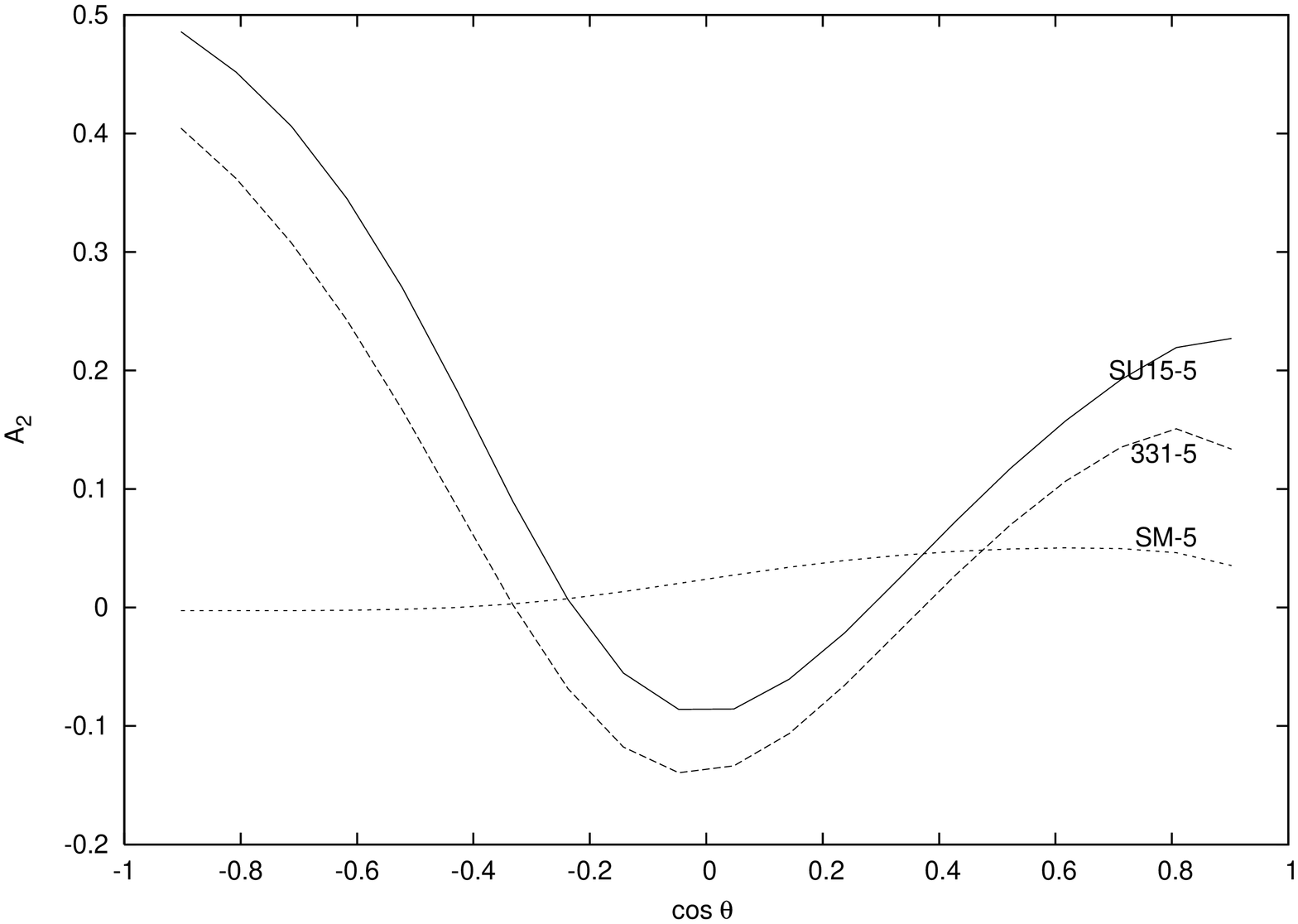}}}
\caption{\label{fig5} Polar angle dependence of spin asymmetry
$A_2(cos\theta)$ at $\sqrt{s} = 500 GeV$; Curve (SM-5) shows the
standard model predictions for $A_2(cos\theta)$, while (331-5)
represents the corresponding expectations for the minimal {331}
model, for a mass $M_Y = 1.2 TeV$. Curve (SU15-5) shows the prediction
for the SU(15) model.}
\end{figure}

\par

\begin{figure}
\rotatebox{-90}{\scalebox{0.35}{\includegraphics{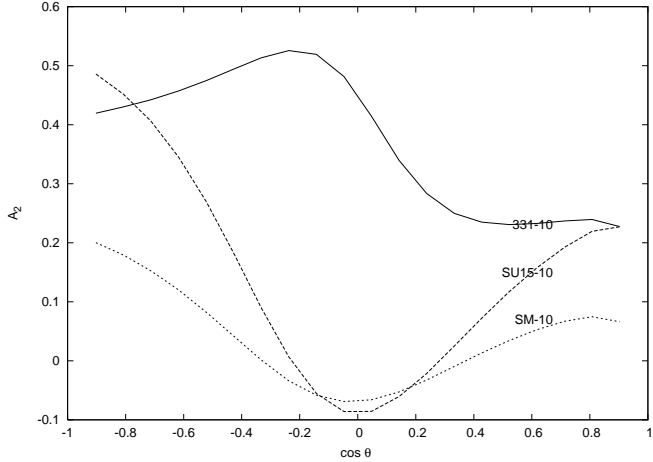}}}
\caption{\label{fig6} Polar angle dependence of spin asymmetry
$A_2(cos\theta)$ at $\sqrt{s} = 1 TeV$; Curve (SM-10) shows the
standard model predictions for $A_2(cos\theta)$, while (331-10)
represents the corresponding expectations for the minimal {331}
model, for a mass $M_Y = 1.2 TeV$. Curve (SU15-10) shows the prediction
for the SU(15) model.}
\end{figure}

In the definitions of the three asymmetries above, $a$ and $b$ refer to the electron and positron beams respectively. Fig.\ref{fig4} shows the angular dependence of asymmetry $A_1$ for a bileptonic mass $M_Y = 1.2 \, TeV$, at $\sqrt{s} = 1 TeV$. We do not show the corresponding $500$ $GeV$ curve, since the $A_1$ sensitivity is considerably lower at this energy. Asymmetry $A_2$ has proved to be quite sensitive to the presence of vector bileptons. For $\sqrt{s} = 500 GeV$ this is shown in Fig.\ref{fig5}, where the $A_2$ dependence on $\cos{\theta}$ is seen to be weaker in the SM, than in the two other models endowed with bileptons. The discriminating power of $A_2$ is even more evident at $\sqrt{s} = 1 TeV$, as depicted in Fig.\ref{fig6}. The resulting errors associated with the calculation of longitudinal asymmetries $A_1$ and $A_2$ were small, and hence the error bars are not shown in the figures. In particular, using bootstrap resampling techniques \cite{BOOT} we confirmed
 that statistical errors are indeed small. Some of the systematic errors, such as the uncertainties on the degrees of beam polarization, were directly included in the simulation. As a matter of fact, we expect some of these systematic effects to cancel out in the polarization asymmetries.

\begin{figure}
\rotatebox{-90}{\scalebox{0.35}{\includegraphics{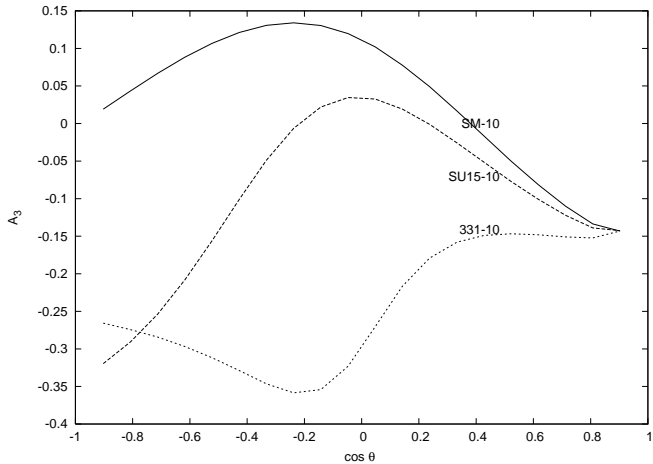}}}
\caption{\label{fig7} Polar angle dependence of spin asymmetry
$A_3(cos\theta)$; Curve (SM-10) shows the standard model predictions
for $A_3(cos\theta)$ at $\sqrt{s} = 1 TeV$, while (331-10) represents
the corresponding expectations for the minimal {331} model, for a
mass $M_Y = 1.2 TeV$. Curve (SU15-10) shows the prediction for the
SU(15) model.}
\end{figure}

Asymmetry $A_3$ concerns the difference between a polarized distribution and its unpolarized counterpart, and is
displayed in Figs.\ref{fig7} and \ref{fig8} for both $1 \, TeV$ and $500 \, GeV$ respectively. For both energies its the $331$ minimal model to give the most different predictions compared with the SM. A curious fact that must be noted it's the change of behavior that only the $331$ model suffers at $1 \, TeV$. At $500 \, GeV$ the three curves have qualitative similar distributions, but at  $1 \, TeV$, although this similarity holds for the SM and the $SU(15)$ model, this is no longer valid for the $331$ model, as one can see in Fig.\ref{fig7}.

Finally, we analyze the discovery potential of the spin asymmetries $A_{1T}$ and $A_{2T}$ for Bhabha scattering. We define:

\begin{figure}
\rotatebox{-90}{\scalebox{0.35}{\includegraphics{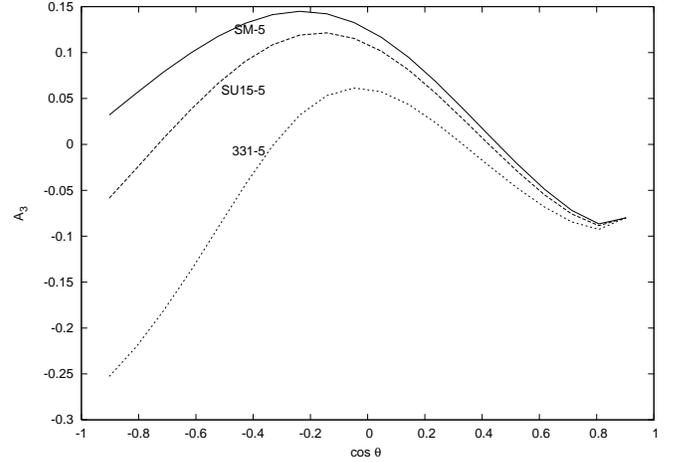}}}
\caption{\label{fig8} Polar angle dependence of spin asymmetry
$A_3(cos\theta)$; Curve (SM-5) shows the standard model predictions
for $A_3(cos\theta)$ at $\sqrt{s} = 500 GeV$, while (331-5) represents
the corresponding expectations for the minimal {331} model, for a
mass $M_Y = 1.2 TeV$. Curve (SU15-5) shows the prediction for the
SU(15) model.}
\end{figure}

\begin{figure}
\rotatebox{-90}{\scalebox{0.35}{\includegraphics{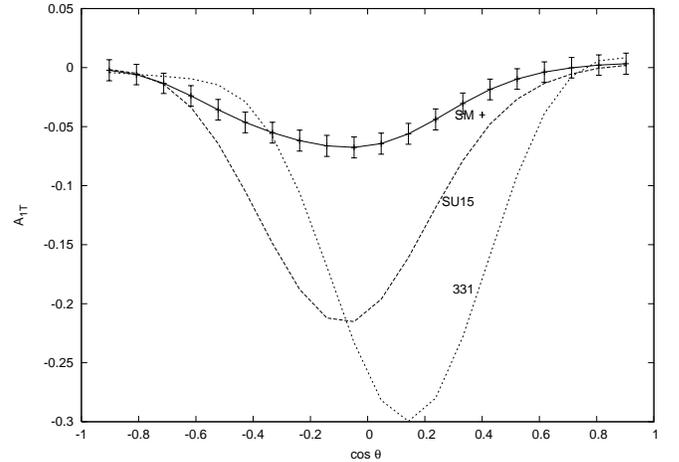}}}
\caption{\label{fig9} Transverse polarization asymmetry
$A_{1T}(cos\theta)$ as a function of $cos\theta$, at $\sqrt{s} = 1
TeV$. The line with error bars (SM) corresponds to the standard model prediction, while the other curves represent the asymmetry for the minimal $331$ model and $SU(15)$ model with $M_Y = 800 GeV$. The error bars for (SU15) and (331) are similar to those of the SM on a bin-to-bin basis.}
\end{figure}

\begin{equation}
A_{1T}(cos\theta) = \frac{ \int_{(+)} d\phi \, {d^2\sigma \over
d(cos\theta)d\phi} - \int_{(-)} d\phi \, {d^2\sigma \over
d(cos\theta)d\phi}} { \int_{(+)} d\phi \, {d^2\sigma \over
d(cos\theta)d\phi} + \int_{(-)} d\phi \, {d^2\sigma \over
d(cos\theta)d\phi}} \qquad ,
\end{equation}
where the subscript $+(-)$ indicates that the integration over the
azimuthal angle $\phi$ is to be carried out over the region of
phase space where $cos \, 2\phi$ is positive (negative).

Fig.\ref{fig9} shows the result for $A_{1T}$ considering $M_Y = 800 \,  GeV$ and energy of $1 \, TeV$. Although the errors are bigger than the case with longitudinal polarization, the observable allows a clear vector bilepton signal for the considered extended models. The changes in the shapes of the curves for $A_{1T}(cos\theta)$ for the two nonstandard models are small when the bilepton mass is increased to $1.2 \,TeV$.

\begin{figure}
\rotatebox{-90}{\scalebox{0.35}{\includegraphics{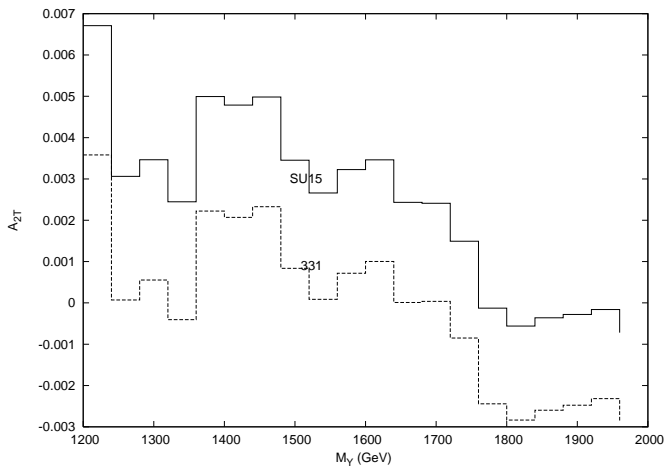}}}
\caption{\label{fig10} Transverse polarization asymmetry $A_{2T}$
as a function of $M_Y$ at $1 TeV$. The solid histogram (331) refers
to the minimal $331$ model. The resulting curve for the $SU(15)$
model (SU15) is also shown for comparison.}
\end{figure}

The integrated version of the asymmetry $A_{2T}$, were also investigated
\begin{equation}
A_{2T} = \frac{ \int_{(+)} d(cos\theta) d\phi \, {d^2\sigma \over
d(cos\theta)d\phi} - \int_{(-)} d(cos\theta)d\phi \, {d^2\sigma
\over d(cos\theta)d\phi}}{ \int_{(+)} d(cos\theta)d\phi \,
{d^2\sigma \over d(cos\theta)d\phi} + \int_{(-)} d(cos\theta)d\phi
\, {d^2\sigma \over d(cos\theta)d\phi}} \qquad ,
\end{equation}
where the integrations are consistent with the cuts specified in
section III.
The result is shown in Fig.\ref{fig10}, where the mass dependence of $A_{2T}$ for $\sqrt{s} = 1
TeV$ for both the $SU(15)$ and the $331$ model were plotted. Qualitatively, there are no differences between the curves. Nevertheless the absolute value of $A_{2T}$ is greater for the $SU(15)$ model. For the sake of comparison, the approximate value of $A_{2T}$ for the standard model was found to be $-0.1$ .

\section{Bounds}

\begin{figure}
\rotatebox{-90}{\scalebox{0.35}{\includegraphics{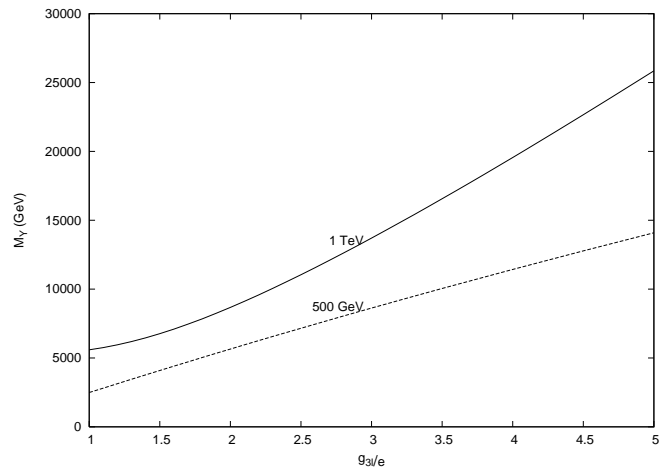}}}
\caption{\label{fig11} $95\%$ C. L. contour plots on the
$(g_{3l}/e,M_Y)$ plane for longitudinally polarized Bhabha
scattering, at the NLC center-of-mass energies $\sqrt{s} = 500
GeV$ (lower curve) and $\sqrt{s} = 1 TeV$ (upper curve).}
\end{figure}

\begin{figure}
\rotatebox{-90}{\scalebox{0.35}{\includegraphics{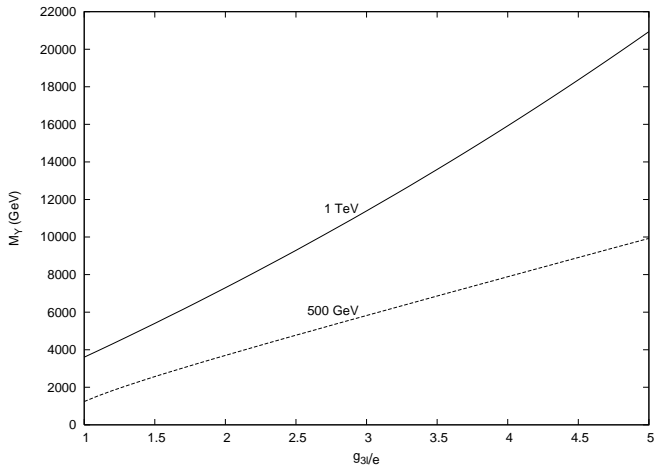}}}
\caption{\label{fig12} $95\%$ C. L. contour plots on the
$(g_{3l}/e,M_Y)$ plane for unpolarized Bhabha scattering, at the NLC
center-of-mass energies $\sqrt{s} = 500 GeV$ (lower curve) and
$\sqrt{s} = 1 TeV$ (upper curve).}
\end{figure}

We estimated lower bounds on the bilepton mass by a $\chi^2$ test, comparing the angular
distribution $d\sigma/d(cos\theta)$ of the final-state electrons, modified by
the presence of a vector bilepton, with the corresponding standard model distribution. Here only longitudinal beam polarization was considered. We assumed that the experimental data will be
well described by the standard model predictions. We treat $g_{3l}$ and $M_Y$ as free parameters in an effective theory and define our $\chi^2$ estimator as
\begin{equation}
\chi^2(g_{3l},M_Y) = \sum_{i=1}^{N_b} {\biggl( {N_i^{SM}- N_i
\over \Delta N_i^{SM}}\biggr)^2}
\end{equation}
where $N_i^{SM}$ is the number of standard model events detected
in the $i^{th}$ bin, $N_i$ is the number of events in the $i^{th}$
bin as predicted by the model with bileptons, and $\Delta N_i^{SM}
= \sqrt{(\sqrt {N_i^{SM}})^2 + (N_i^{SM}\epsilon)^2}$ the
corresponding total error, which combines in quadrature the
Poisson-distributed statistical error with the systematic error, which for the latter we assumed to be of $\epsilon = 5\%$ for each measurement. The angular range $\vert cos\theta \vert <
0.95$ was divided into $N_b= 20$ equal-width bins. For fixed values of the coupling
$g_{3l}$ the bilepton mass $M_Y$ was varied as a free parameter to determine the $\chi^2$
distribution. The $95\%$ confidence level bound corresponds to an
increase of the $\chi^2$ by $3.84$ with respect to the minimum
$\chi^2_{min}$ of the distribution. Fig.\ref{fig11} presents the resulting $95\%$ C. L. contour plots on the $(g_{3l}/e,M_{Y})$ plane for the
nominal center-of-mass energies $\sqrt{s} = 500 GeV$ and $\sqrt{s}
= 1 TeV$. The unpolarized case is shown in Fig.\ref{fig12}.
We also calculated the corresponding $95\%$ C. L. limits on the
bilepton mass at these ILC energies, considering only the
minimal $331$ model. The results are displayed in Table I. 

\section{Conclusions}

We found that the analysis of spin asymmetries in Bhabha scattering leads to tighter bounds than the counterparts obtained from $M\o ller$ scattering \cite{MEIROSE}. We emphasize that several beam effects such as ISR,  beamstrahlung, beam energy and polarization spreads were taken into effect in our simulations. 
In brief, we showed that polarization-dependent observables may be used to search for signals of vector bileptons in Bhabha scattering, at the next generation of linear colliders. Should vector bileptons be detected, the underlying nonstandard model describing their interactions could be determined by the discriminating power of combined measurements of spin asymmetries. Otherwise, polarized Bhabha scattering would still allow that useful experimental bounds be set on the bilepton mass and couplings.

\begin{table}
\caption{\label{table1}$95\%$ C. L. limits on the bilepton mass
in the minimal $331$ model, at NLC energies}
\begin{ruledtabular}
\begin{tabular}{lcr}
Polarization&$\sqrt{s}$=500 GeV&$\sqrt{s}$=1 \, TeV \\
\hline
unpolarized & 4485 GeV & 9519 GeV\\
polarized & 5978 GeV & 11056 GeV\\
\end{tabular}
\end{ruledtabular}
\end{table}


\begin{acknowledgments}
 This work was supported by CNPq.
\end{acknowledgments}


\begin{thebibliography}{99}
\bibitem{CUYDAV} F. Cuypers and S. Davidson, Eur. Phy. J. C {\bf 2},503
(1998) [hep-ph/9609487] and references therein.
\bibitem{FRAPLE} F. Pisano and V. Pleitez, Phys. Rev. D {\bf 46}, 410
(1992); P. H. Frampton, Phys. Rev. Lett. {\bf 69}, 2889 (1992).
\bibitem{SUFI} P. H. Frampton and B.-H. Lee, Phys. Rev. Lett. {\bf 64}, 619
(1990); P. H. Frampton and T.W. Kephart, Phys. Rev. D{\bf 42}, 3892 (1990); P. H. Frampton, Int. J. Mod. Phys. A{\bf 15}, 2455 (2000).
\bibitem{PPT} F. Pisano, V. Pleitez and M. D. Tonasse, IFT-UNESP
preprint IFT-P.043/97 and references therein.
\bibitem{TUL} M. B. Tully and G. C. Joshi, Phys. Lett. B466, 333 (1993); hep-ph/9905552.
\bibitem{WILL} Willmann et. al., Phys. Rev. Lett. 82, 49 (1999).
\bibitem{GUTI} N. Gutierrez, R. Martinez and F. Ochoa, arXiv:0802.0310v1 [hep-ph].
\bibitem{DIAS1} A. G. Dias, R. Martinez, V. Pleitez, Eur. Phys. J. C39, 101 (2005)
\bibitem{DIAS2} Alex Gomes Dias, Phys. Rev. D71, 015009 (2005).
\bibitem{SKR} M. Skrzypek and S. Jadach, Z. Phys. C{\bf 49}, 577 (1991).
\bibitem{BDH} M. B\"{o}hm, Ansgar Denner and W. Hollik, Nucl. Phys. B304, 687 (1988).
\bibitem{CLIC} The CLIC Study Team, "A $3 TeV$ $e^+e^-$ Linear Collider
Based on CLIC Technology", report CERN 2000-008 (2000).
\bibitem{PESK} M. Peskin, SLAC-TN-04-032 (1999).
\bibitem{BEAM} Beam Parameters and Lattices for NLC2003,\\ http://www-project.slac.stanford.edu/lc/local/\\documentation/pdf/NLC2003$\_$config.pdf.
\bibitem{FRIZ} Paul H. Frampton and Daniel Ng, Phys. Rev. D {\bf 45}, 4240
(1992); Thomas Rizzo, Phys. Rev. D {\bf 46}, 910 (1992).
\bibitem{OLS} Haakon A. Olsen and Per Osland, Phys. Rev. D{\bf 25}, 2895
(1982).
\bibitem{SMEAR} R. Settles, H. Spiesberger and W. Wiedenmann, Smear
version 3.02; http://www.desy.de/hspiesb/smear.html.
\bibitem{MOOR}  G. Moortgat-Pick and H. Steiner, Eur. Phys. J. direct C{\bf 6} (2001) 1, hep-ph/0106155.
\bibitem{BOOT}  B. Efron and R.J. Tibshirani, An Introduction to Bootstrap, ed. Chapman and Hall (1993).
\bibitem{MEIROSE} B. Meirose and A.J. Ramalho, Phys. Rev. D {\bf 73}, 075013 (2006).




\end{thebibliography}

\end{document}